\documentstyle{elsart}
\begin{document}
%\rightline{LA-UR-96-2031}
\begin{frontmatter}
\title{About the Interpretation of Gravitationally Induced Neutrino
Oscillation Phases}

\author{D. V. Ahluwalia}
\footnote{E-mail address: ahluwalia@nus.lampf.lanl.gov .}
\address{Mail Stop H--846, Physics (P--25) and Theory (T--5) Division \\ 
Los Alamos National Laboratory, Los Alamos, NM 87545, USA}

\author{C. Burgard}
\footnote{E-mail address: christoph.burgard@cern.ch .}
\address{
Universit\"at Hamburg/DESY,
II. Institut f\"ur Experimentalphysik\\
Notkestr.85,
D-22607 Hamburg, Germany}

\begin{abstract}
We present some thoughts on how to interpret the gravitionally induced neutrino 
oscillation phases presented by us in our 1996 Gravity Research Foundation Essay. 
\end{abstract}

\end{frontmatter}

In a recent paper we discussed the modification of the neutrino oscillation phases
due to the presence of gravity \cite{grf96}. In a comment by Bhattacharya, 
Habib and Mottola (BHM) our results were rederived
and an interpretation of them were given \cite{BHM}. Here we would like
to present our own interpretation, which stresses less the general
relativistic aspect of the results but is more interested in the
way gravity and quantum mechanics intermingle with each
other \cite{grf96,grf94}.
In particular our essay studied the detailed interplay of
the quantum mechanical principle of the linear superposition
and general relativity's principle of equivalence. From a purely general 
relativistic point of view, both the considerations of our essay and the 
physics of neutron interferometry experiments \cite{COW} may
have nothing exciting \cite{BHM}, but it must be emphasised how
the interplay of the principle of superposition and the presence
of gravity produces the gravitationally induced neutrino oscillation phases
in a manner that ensures compliance with the principle of equivalence.

First, begin with classical and quantum mechanical considerations
for a  single mass eigenstate. For a single mass eigenstate the 
classical effects of gravitation may be considered to depend
on  a force, $\vec{F}$, while the quantum--mechanical 
effects are determined by the gravitational 
interaction energy.
The gravitational 
interaction energy for  non-relativistic particle of mass $m$ has the 
same form as in Newtonian theory and reads 
$U_{int.} = m \times \phi$, while for a relativistic
particle (as shown below) it is $U_{int.}= (E/c^2)\times \phi$.
The $\phi = - GM/r$ represents the gravitational potential in the weak field
limit for an object of mass, $M$, in the standard notation \cite{SW};
with $\vec{F} = - \vec{\nabla} \phi$. 
Along an eqi--$\phi$ surface 
the $\vec{F}$ vanishes and there are no classical effects in this direction.
The constant potential along a segment of a  eqi--potential surface  
can be removed by going to an appropriately accelerated fame.

These are well known text--book statements \cite{SW,JJS,LBO}.
However, we now explicitly note what is not always fully appreciated.
Quantum mechanically, the mass eigenstate, assumed to have
remained stationary at a given spacial position,  picks up a global phase factor
$\exp\left(- i\,  m \,\phi\, t/\hbar \right)$. Again, there are no
physical consequences. 
If we now consider a physical state that is in 
a  linear superposition
of different mass eigenstates then 
physically observable 
relative phases are induced between various mass 
eigenstates.
\footnote{The observability of these phases is not for  a local observer,
but for an observer making measurements stationed at a different eqi-$\phi$
surface.}
 Specifically, 
on an eqi--$\phi$ surface the gravitational force $\vec{F}$ vanishes,
while the relative quantum mechanical phases induced in the evolution
of a linear superposition of mass eigenstates do not. 

For the  Newtonian case considered: 
\begin{eqnarray}
&&\mbox{
The gravitationally induced (time--)oscillatory
phase}   \nonumber \\ 
&&\quad=\,\Phi \times \,\,\mbox{The (time--)oscillatory  phase without gravity}\quad,
\end{eqnarray}
 with 
$\Phi=\phi/c^2$,
the dimensionless gravitational potential.
However, as one cannot measure a gravitational potential by a local
measurement,  one
 needs to make similar observations on two different eqi--$\phi$ surfaces
to discern the presence of gravity. 
The gravitationally induced oscillatory phase, denoted by
$\varphi^G_{\jmath\imath}$ 
for neutrino oscillations in \cite{grf96}, is
an addition to the oscillatory phase without 
gravity (denoted by
$\varphi^0_{\jmath\imath}$ 
for neutrino oscillations in \cite{grf96}). Since neutrino--oscillation
experiments are only sensitive to the sum of both phases, one needs to make
similar observations on two different eqi--$\phi$ surfaces and compare them in
order to measure the presence of the gravitationally induced phase.
One of these surfaces should be the surface at spacial infinity 
to extract the full gravitationally induced phase. Alternately, one may wish 
to  compare one's results with an experiment performed in a freely falling 
orbiter around the massive object.
Specifically, the
sense in which this comparison is to be performed is identical to that in
which one measures a gravitational red shift of stellar spectra on Earth.
The quantum mechanically created clock, via the time-oscillation of the
mass eigenstates in the linear superposition, suffers the gravitational
red shift as demanded by general relativity when the gravitationally
induced oscillatory phase is taken into account.

In our 1996 Gravity Research Foundation Essay \cite{grf96}
the above noted non-relativistic
observations were  appropriately modified and applied 
to neutrino weak flavor eigenstates which are empirically 
indicated  to be linear superposition of mass eigenstates.
We confirmed the demands of general relativity in a quantum
context.

BHM have shown in their communication \cite{BHM}
that by a {\em local} measurement one cannot measure
a {\em local} gravitational potential. We agree.

In reference to BHM's comment \cite{BHM} one may wish to note:

\begin{enumerate}

\item
Exactly how the 
relativistic expression for $U_{int}$,
mentioned above, is obtained.
In the weak field limit the force on a mass eigenstate, of mass $m$,
in the Schwarzschild gravitational environment mass of  M reads \cite{LBO}
\begin{equation}\vec{F} = -\, {{G M m \gamma}\over {r^3}}\left[
\left(1+\beta^2\right)\vec{r}-\left(\vec{r}\cdot\vec{\beta}\right)
\vec{\beta}\right]\quad.
\end{equation}
In this equation, $
\beta=\vec{v}/{\vert\vec{v}\vert} $, with
$\vec{v}$ the velocity of mass eigenstate, and $\gamma = \left(1-\beta^2\right)
^{-1/2}$. Assuming the mass eigenstate to be relativistic and setting
$m\gamma=E/c^2$, we have
\begin{equation}
U_{int.} = \int_\infty^r \vec{F}\cdot \mbox{d}\vec{r^\prime}
=-\, {{GME}\over {r\,c^2}}= - (E/c^2)\times \phi\quad.
\end{equation}

\item
Theoretically, the prediction of general relativity as regards any clock
(classically driven, or quantum mechanically, 
with non-relativistic mechanism or relativistic) and the prediction arising
from quantum evolution that incorporates gravity via an interaction
energy term (in a manner similar to the classic neutron interferometer
experiments and their analysis \cite{COW,COWd}) are in mutual agreement.
However, 
recently it has been suggested that the atmospheric and solar
neutrino data could be explained by a violation of the equivalence
principle \cite{HLP}
 and it is, therefore, an important matter to understand
in detail how the quantum mechanics and gravity work together for neutrino 
oscillations. Apart from this motivation, the problem is of interest in its
own right to understand explicitly, and in detail, how the principle
of equivalence and the principle of linear superposition of quantum mechanics
intermingle \cite{grf94}.

\item
In reference to the discussion following Eq. (10) of the comment \cite{BHM} by
BHM one needs to note that it is not
necessary (even in a semi-classical framework), or even correct,
to note (as the authors of the comment \cite{BHM}  under consideration do)
``Since the energy is fixed but
the masses are different, if interference is to be observed at the
same final spacetime point $(r_B,t_B)$, the relevant components of the
wave function could not both have started from the same initial
spacetime point $(r_A,t_A)$.''
It needs to be appreciated that a wave function, after having 
evolved over a certain distance,  may develop more than one
spacial peak (perhaps corresponding to each of the mass eigenstate)
and yet ``collapse'' to a single space--time point (or, appropriately
defined spacial region governed by the uncertainty principle) in  a
weak flavor measurement. This, we believe, is the orthodox
text--book wisdom \cite{tg,HL,KPbook} and we see no need to violate \cite{BHM,sh} 
it in neutrino oscillation phenomenology. 
\end{enumerate}

We intend to take up
the subject matter in more detail elsewhere and show how to extend the 
conceptual framework to terrestrial experiments (with atomic systems
and superconducting devices)
where the weaker gravity of the Earth can be compensated by
integrating the gravitationally induced effects over time.

In summary, and in reference to the BHM comment \cite{BHM}, 
we stand by our conclusions presented in  \cite{grf96}.

{\bf Acknowledgements}
We wish to extend our thanks to
Terry Goldman, Yuval Grossman, David Ring and
Svend Rugh for extended discussion on the subject.
We also thank Steven Weinberg for a comment on the essay
that led to a deeper
understanding of the interplay of gravitation and quantum mechanics,
and Sam Werner for providing us with a copy of Ref. \cite{COWd}
before its publication. The acknowledged do not necessarily agree, or
disagree, with the manuscript.

{\em 
This work was done, in part, under the auspices of the 
U. S. Department of Energy; and supported by facilities and
financial support of the groups P-25 and T-5
of the Los Alamos National Laboratory.}


\begin{thebibliography}{999}
\bibitem{sh}
In order to prevent confusion arising when comparing the results
of BHM's communication \cite{BHM} and our essay \cite{grf96}, let us stress that we
defined the  four-momentum $p^\mu$ as the one from special relativity and
treated the difference between $g^{\mu\nu}$ and the flat metric
$\eta^{\mu\nu}$ as perturbation on top of the flat metric $\eta^{\mu\nu}$.
Thus the symbols $p^\mu$, $E$ and $\vec p$ in \cite{grf96} 
and the corresponding symbols in
\cite{BHM} do not describe the same
objects.  
Within our framework 
$
\left[E -\left({G M E}/{c^2 r}\right)\right]
$ is conserved and our results are obtained 
to first order in $\Delta m^2$ and $\Phi$.
The derivation of the  kinematic and the 
gravitationally induced oscillatory phases 
starting from Eq. (4) of \cite{grf96}
is a straightforward exercise (with the algebraic results also
rederived by BHM in \cite{BHM})
to any one familiar with the 
standard neutrino oscillation
phenomenology and was therefore skipped in \cite{grf96}.
\end{thebibliography}
\end{document}